\documentclass[10pt,twoside]{article}
\usepackage{amssymb}
\usepackage{amsmath}
\usepackage[T1]{fontenc}
\usepackage[latin9]{inputenc}
\usepackage{amsfonts}
\usepackage{float}
\usepackage{graphicx}
\usepackage{lscape}
\usepackage{url}
\usepackage{fancyhdr}
\usepackage{fancybox}
\usepackage[linktocpage,colorlinks=true,linkcolor=black,filecolor=black,urlcolor=black]{hyperref}
\usepackage{multicol}
\usepackage{nicefrac}

\usepackage{algorithm}
\usepackage{algpseudocode}

\def\tableskip{\vskip 10pt plus 2pt minus 2pt\relax}
\def\figureskip{\vskip 10pt plus 2pt minus 2pt\relax}
\def\limfunc#1{\mathop{\rm #1}}
\def\func#1{\mathop{\rm #1}\nolimits}
\newlength{\figurewidth}
\newlength{\figureheight}
\graphicspath{{Figures/}}
\def\limfunc#1{\mathop{\rm #1}}
\def\func#1{\mathop{\rm #1}\nolimits}

\newcommand{\bN}{\mathrel{\;\;\,}}

\begin{document}

\title{\textbf{Momentum Strategies with $L_1$ Filter}}
\author{
Tung-Lam Dao\\
Capital Fund Management, Paris\\
\texttt{tung-lam.dao@cfm.fr}
}
\date{\today}

\maketitle

\begin{abstract}
In this article, we discuss various implementation of $L_1$
filtering in order to detect some properties of noisy signals. This
filter consists of using a $L_1$ penalty condition in order to
obtain the filtered signal composed by a set of straight trends or
steps. This penalty condition, which determines the number of
breaks, is implemented in a constrained least square problem and is
represented by a regularization parameter $\lambda$ which is
estimated by a cross-validation procedure. Financial time series are
usually characterized by a long-term trend (called the global trend)
and some short-term trends (which are named local trends). A
combination of these two time scales can form a simple model
describing the process of a global trend process with some
mean-reverting properties. Explicit applications to momentum
strategies are also discussed in detail with appropriate uses of the
trend configurations.
\end{abstract}

\medskip


\section{Introduction}

Trend detection is a major task of time series analysis from both
mathematical and financial point of view. The trend of a time series
is considered as the component containing the global change which is
in contrast to the local change due to the noise. The procedure of
trend filtering concerns not only the problem of denoising but it
must take into account also the dynamic of the underlying process.
That explains why mathematical approaches to trend extraction have a
long history and this subject still gives a great interest in the
scientific community \footnote{For a general review, see Alexandrov
\textsl{et al.} (2008).}. In an investment perspective, trend
filtering is the core of most momentum strategies developed in the
asset management industry and the hedge funds community in order to
improve performance and to limit risk of portfolios.\bigskip

The paper is organized as follows. In section 2, we discuss the
trend-cycle decomposition of time series and review general
properties of $L_1$ and $L_2$ filtering. In section 3, we describe
the $L_1$ filter with its various extensions and the calibration
procedure. In section 4, we apply $L_1$ filters to some momentum
strategies and present the results of some backtests with the S\&P
500 index. In section 5, we discuss the possible extension to the
multivariate case and we conclude in the last section.

\section{Motivations}

In economics, the trend-cycle decomposition plays an important role
to describe a non-stationary time series into permanent and
transitory stochastic components. Generally, the permanent component
is assimilated to a trend whereas the transitory component may be a
noise or a stochastic cycle. Moreover, the literature on business
cycle has produced a large number of empirical research on this
topic (see for example Cleveland and Tiao (1976), Beveridge and
Nelson (1991), Harvey (1991) or Hodrick and Prescott (1997)). These
last authors have then introduced a new method to estimate the trend
of long-run GDP. The method widely used by economists is based on
$L_2$ filtering. Recently, Kim \textsl{et al.} (2009) have developed
a similar filter by replacing the $L_2$ penalty function by a $L_1$
penalty function.\bigskip

Let us consider a time series $y_t$ which can be decomposed by a
slowly varying trend $x_t$ and a rapidly varying noise
$\varepsilon_t$ process:
\begin{equation*}
y_t = x_t + \varepsilon_t
\end{equation*}
Let us first remind the well-known $L_2$ filter (so-called
Hodrick-Prescott filter). This scheme consists to determine the
trend $x_t$ by minimizing the following objective function:
\begin{equation*}
\frac{1}{2}\sum_{t=1}^n \left(y_t-x_t\right)^2 +
\lambda\sum_{t=2}^{n-1}\left(x_{t-1}-2x_{t}+x_{t+1}\right)^2
\end{equation*}
with $\lambda >0$ the regularization parameter which control the
competition between the smoothness of $x_t$ and the residual
$y_t-x_t$ (or the noise $\varepsilon_t$). We remark that the second
term is the discrete derivative of the trend $x_t$ which
characterizes the smoothness of the curve. Minimizing this objective
function gives a solution which is the trade-off between the data
and the smoothness of its curvature. In finance, this scheme does
not give a clear signature of the market tendency. By contrast, if
we replace the $L_2$ norm by the $L_1$ norm in the objective
function, we can obtain more interesting properties. Therefore, Kim
\textsl{et al.} (2009) propose to consider the following objective
function:
\begin{equation*}
\frac{1}{2}\sum_{t=1}^n \left(y_t-x_t\right)^2+\lambda\sum_{t=2}^{n-1}\left|x_{t-1}-2x_{t}+x_{t+1}\right|
\end{equation*}
This problem is closely related to the Lasso regression of
Tibshirani (1996) or the $L_1$ regularized least square problem of
Daubechies \textsl{et al.} (2004). Here, the fact of taking the
$L_1$ norm will impose the condition that the second derivation of
the filtered signal must be zero.
Hence, the filtered signal is composed by a set of straight trends and breaks\footnote{A break is the position where the trend of signal changes.}.
The competition between these two terms in the objective
function turns to the competition between the number of straight trends (or number of breaks) and
the closeness to the raw data. Therefore, the smoothing parameter
$\lambda$ plays an important role for detecting the number of
breaks. In the later, we present briefly how the $L_1$ filter works
for the trend detection and its extension to mean-reverting
processes. The calibration procedure for $\lambda$ parameter will be
also discussed in detail.

\clearpage

\section{$L_1$ filtering schemes}

\subsection{Application to trend-stationary process}
The Hodrick-Prescott scheme discussed in last section can be
rewritten in the vectorial space $\mathbb{R}^n$ and its $L_2$ norm
$\left\| \cdot \right\|_2$ as:
\begin{equation*}
\frac{1}{2}\left\|y-x\right\|_2^2+\lambda\left\|Dx\right\|_2^2
\end{equation*}
where $y=\left(y_1,\dots,y_n\right)$,
$x=\left(x_1,\dots,x_n\right)\in \mathbb{R}^n$ and the $D$ operator
is the $\left(n-2\right)\times n$ matrix:
\begin{equation}
D = \left[
\begin{array}{rrrrrrrr}
1 & -2 & 1  &         &     &    &   \\
  &  1 & -2 & 1       &     &    &   \\
  &    &    &  \ddots &     &    &   \\
  &    &    &       1 &  -2 &  1 &   \\
  &    &    &         &   1 &  2 & 1
\end{array}
\right]
\label{eq:D2}
\end{equation}
The exact solution of this estimation is given by
\begin{equation*}
x^{\star}=\left( I +2 \lambda D^{\top}D \right)^{-1}y
\end{equation*}
The explicit expression of $x^\star$ allows a very simple numerical
implementation with sparse matrix. As $L_2$ filter is a linear
filter, the regularization parameter $\lambda$ is calibrated by
comparing to the usual moving-average filter. The detail of the
calibration procedure is given in Appendix \ref{L2Calibration}.
\bigskip

The idea of $L_2$ filter can be generalized to a lager class
so-called $L_p$ filter by using $L_p$ penalty condition instead of
$L_2$ penalty. This generalization is already discussed in the work
of Daubechies \textsl{et al.} (2004) for the linear inverse problem
or in the Lasso regression problem by Tibshirani \textsl{et al.}
(1996). If we consider a $L_1$ filter, the objective function
becomes:
\begin{equation*}
\frac{1}{2}\sum_{t=1}^n \left(y_t-x_t\right)^2+\lambda\sum_{t=2}^{n-1}\left|x_{t-1}-2x_{t}+x_{t+1}\right|
\end{equation*}
which is equivalent to the following vectorial form:
\begin{equation*}
\frac{1}{2}\left\|y-x \right\|_2^2+\lambda\left\|Dx\right\|_1
\end{equation*}
It has been demonstrated in Kim \textsl{et al.} (2009) that the dual
problem of this $L_1$ filter scheme is a quadratic program with some
boundary constraints. The detail of this derivation is shown in
Appendix \ref{appendix-dual-L1-T}. In order to optimize the
numerical computation speed, we follow Kim \textsl{et al.} (2009) by
using a \textquotedblleft primal-dual interior point\textquotedblright\ method (see Appendix
\ref{appendix-interior-point}). In the following, we check the
efficient of this technique on various trend-stationary processes.%
\bigskip

\begin{figure}[tbph]
\centering
\caption{$L_1-T$ filtering versus HP filtering for the model (\ref{eq:model1})}
\label{fig:test-model1-t}
\figureskip
\includegraphics[width = \figurewidth, height = \figureheight]{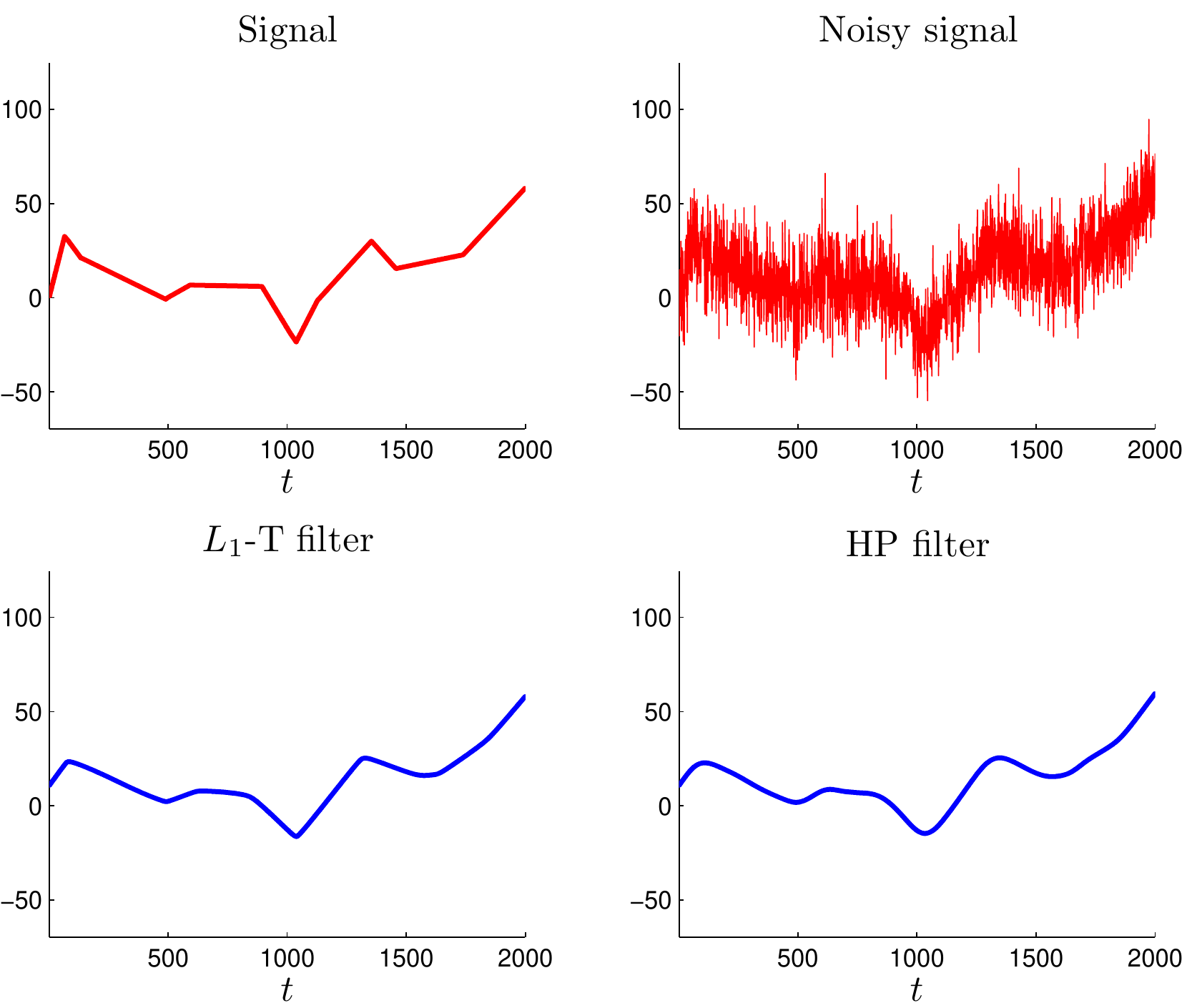}
\end{figure}

\begin{figure}[tbph]
\centering
\caption{$L_1$-T filtering versus HP filtering for the model (\ref{eq:model2})}
\label{fig:test-model2-t}
\figureskip
\includegraphics[width = \figurewidth, height = \figureheight]{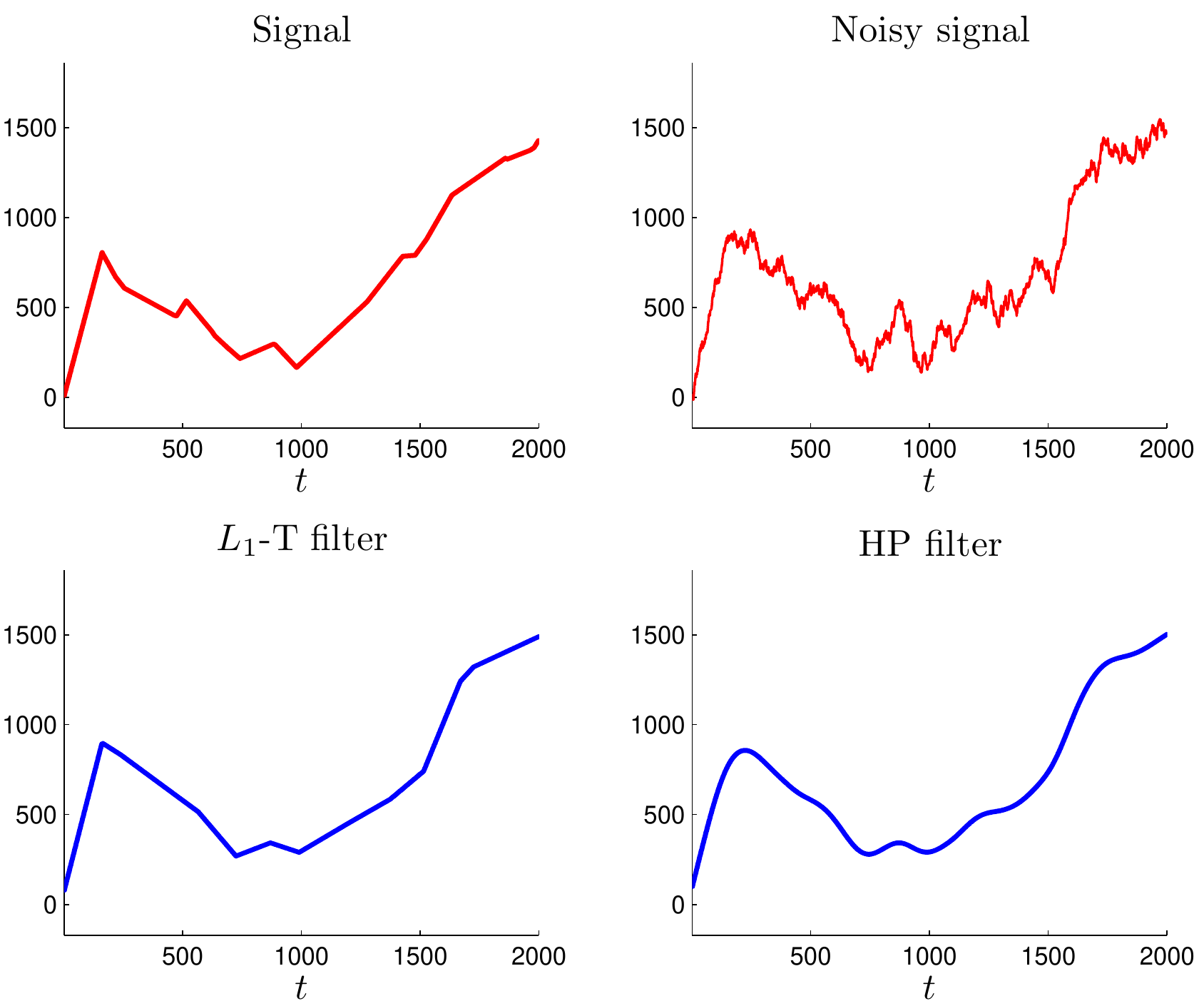}
\end{figure}

The first model consists of data simulated by a set of straight
trend lines with a white noise perturbation:%
\begin{equation}
\left\{
\begin{array}{l}
y_{t}=x_{t}+\varepsilon_{t} \\
\varepsilon_{t}\sim \mathcal{N}\left( 0,\sigma ^{2}\right)  \\
x_{t}=x_{t-1}+v_{t} \\
\Pr \left\{ v_{t}=v_{t-1}\right\} =p \\
\Pr \left\{ v_{t}=b\left( \mathcal{U}_{\left[ 0,1\right] }-\frac{1}{2}%
\right) \right\} =1-p%
\end{array}%
\right.
\label{eq:model1}
\end{equation}%
We present in Figure \ref{fig:test-model1-t} the comparison between
$L_{1}-T$ and HP filtering schemes\footnote{%
We consider $n=2000$ observations. The parameters of the simulation are $%
p=0.99$, $b=0.5$ and $\sigma =15$.}. The top-left graph is
the real trend $x_{t}$ whereas the top-right graph presents
the noisy signal $y_{t}$. The bottom graphs show the results of the
$L_{1}-T$ and HP filters. Here, we have chosen $\lambda=5\,258$ for
the $L_1-T$ filtering and $\lambda=1\,217\,464$ for HP filtering.
This choice of $\lambda$ for $L_1-T$ filtering is based on the
number of breaks in the trend, which is fixed to 10 in this example%
\footnote{We discuss how to obtain $\lambda$ in the next section.}.
The second model model is a random walk generated by the following process:%
\begin{equation}
\left\{
\begin{array}{l}
y_{t} = y_{t-1} + v_{t} + \varepsilon_{t} \\
\varepsilon_{t}\sim \mathcal{N}\left( 0,\sigma ^{2}\right)  \\
\Pr \left\{ v_{t}=v_{t-1}\right\} =p \\
\Pr \left\{ v_{t}=b\left( \mathcal{U}_{\left[ 0,1\right] }-\frac{1}{2}
\right) \right\} =1-p%
\end{array}%
\right.
\label{eq:model2}
\end{equation}%
We present in Figure \ref{fig:test-model2-t} the comparison between
$L_{1}-T$
filtering and HP filtering on this second model\footnote{%
The parameters of the simulation are $p=0.993$, $b=5$ and $\sigma
=15$.}.

\subsection{Extension to mean-reverting process}

\begin{figure}[tbph]
\centering
\caption{$L_1-C$ filtering versus HP filtering for the model (\ref{eq:model3})}
\label{fig:test-model3-c}
\figureskip
\includegraphics[width = \figurewidth, height = \figureheight]{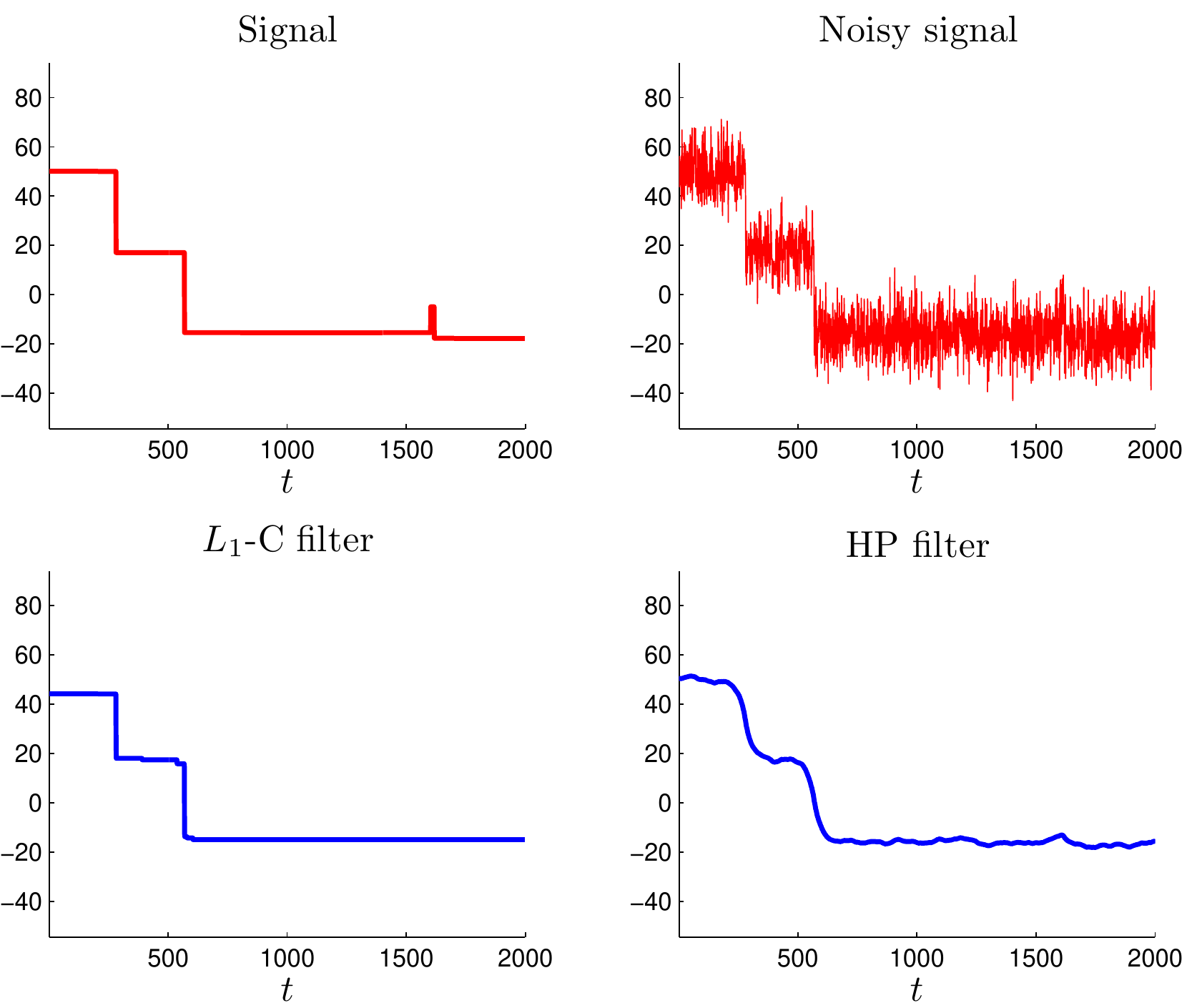}
\end{figure}

\begin{figure}[tbph]
\centering
\caption{$L_1-C$ filtering versus HP filtering for the model (\ref{eq:model4})}
\label{fig:test-model4-c}
\figureskip
\includegraphics[width = \figurewidth, height = \figureheight]{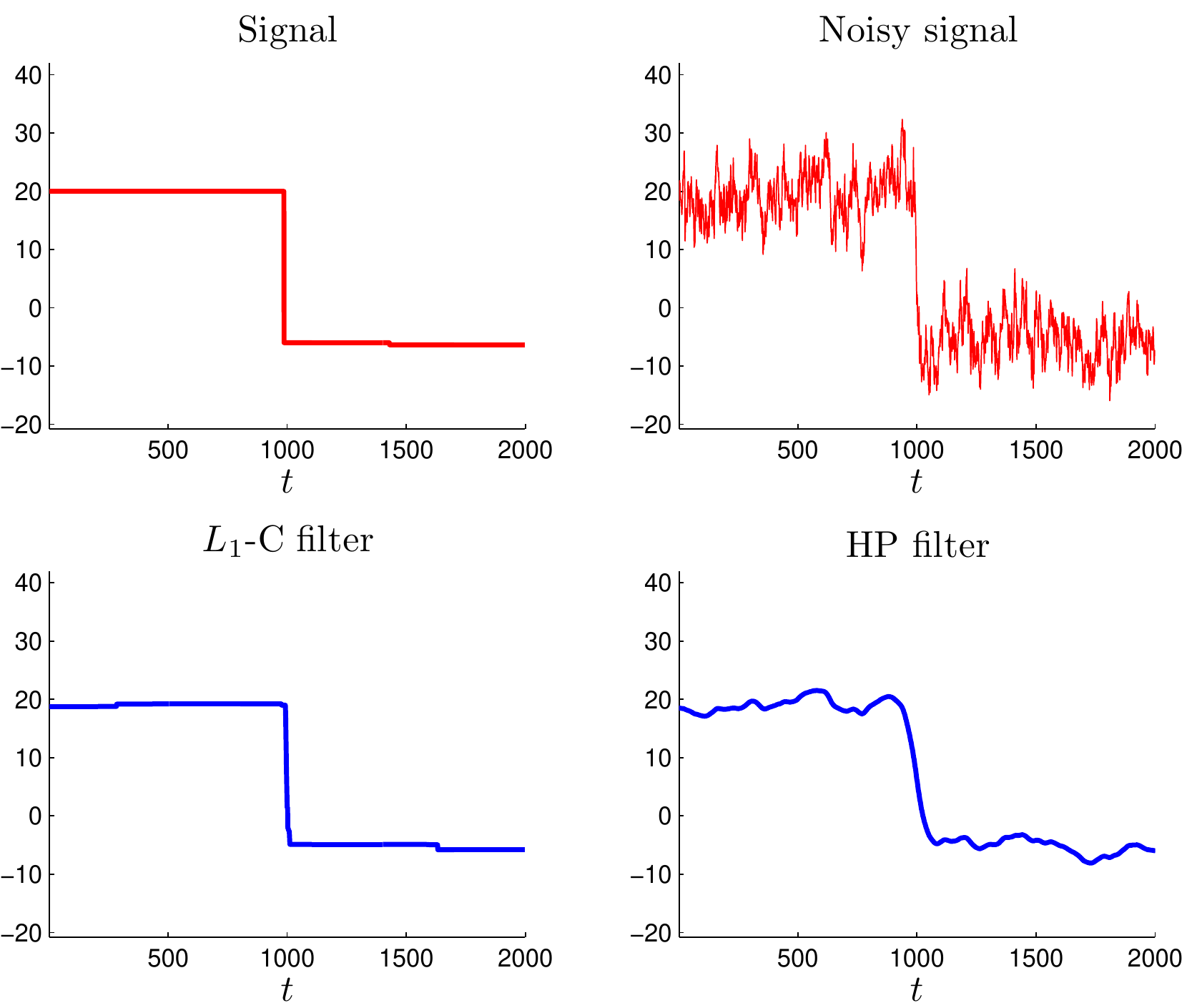}
\end{figure}

As shown in the last paragraph, the use of $L_1$ penalty on the
second derivative gives the correct description of the signal
tendency. Hence, similar idea can be applied for other order of the
derivatives. We present here the extension of this $L_1$ filtering
technique to the case of mean-reverting processes. If we impose now
the $L_1$ penalty condition to the first derivative, we can expect
to get the fitted signal with zero slope. The cost of this penalty
will be proportional to the number of jumps. In this case, we would
like to minimize the following objective function:
\begin{equation*}
\frac{1}{2}\sum_{t=1}^n \left(y_t-x_t\right)^2+\lambda\sum_{t=2}^{n}\left|x_{t}-x_{t-1}\right |
\end{equation*}
or in the vectorial form:
\begin{equation*}
\frac{1}{2}\left\|y-x\right\|_2^2+\lambda\left\|Dx\right\|_1
\end{equation*}
Here the $D$ operator is $\left(n-1\right)\times n$ matrix which is
the discrete version of the first order derivative:
\begin{equation}
D = \left[
\begin{array}{rrrrrrrr}
-1 &  1 & 0 &         &    &    \\
0  & -1 & 1 & 0       &    &    \\
   &    &   &  \ddots &    &    \\
   &    &   &      -1 &  1 &  0 \\
   &    &   &         & -1 &  1
\end{array}
\right]
\label{eq:D1}
\end{equation}
We may apply the same minimization algorithm as previously (see Appendix \ref{appendix-dual-L1-C}).
To illustrate that, we consider the model with step trend lines perturbed by a white noise process:
\begin{equation}
\left\{
\begin{array}{l}
y_{t}=x_{t} + \varepsilon_{t} \\
\varepsilon_{t}\sim \mathcal{N}\left( 0,\sigma ^{2}\right)  \\
\Pr \left\{ x_{t}=x_{t-1}\right\} =p \\
\Pr \left\{ x_{t}=b\left( \mathcal{U}_{\left[ 0,1\right] }-\frac{1}{2}%
\right) \right\} =1-p%
\end{array}%
\right.
\label{eq:model3}
\end{equation}%
We employ this model for testing the $L_1-C$ filtering and HP
filtering adapted to the first derivative%
\footnote{We use the term HP filter in order to keep homogeneous notations.
However, we notice that this filter is indeed the FLS filter proposed by
Kalaba and Tesfatsion (1989) when the exogenous regressors are only a constant.}, which
corresponds to the following optimization program:
\begin{equation*}
\min \frac{1}{2}\sum_{t=1}^{n}\left( y_{t}-x_{t}\right) ^{2}+\lambda
\sum_{t=2}^{n}\left( x_{t}-x_{t-1}\right) ^{2}
\end{equation*}
In Figure \ref{fig:test-model3-c}, we have reported the
corresponding results%
\footnote{The parameters are $p=0.998$, $b=50$ and $\sigma=8$.}. For
the second test, we consider a mean-reverting process (Ornstein-Uhlenbeck process) with mean
value following a regime switching process:
\begin{equation}
\left\{
\begin{array}{l}
y_{t} = y_{t-1} + \theta(x_{t}-y_{t-1}) + \varepsilon_{t} \\
\varepsilon_{t}\sim \mathcal{N}\left( 0,\sigma ^{2}\right)  \\
\Pr \left\{ x_{t}=x_{t-1}\right\} = p \\
\Pr \left\{ x_{t}=b\left( \mathcal{U}_{\left[ 0,1\right] }-\frac{1}{2}\right) \right\} = 1-p%
\end{array}%
\right.
\label{eq:model4}
\end{equation}%
Here, $\mu_t$ is the process which characterizes the mean value and
$\theta$ is inversely proportional to the return time to the mean
value. In Figure~\ref{fig:test-model4-c}, we show how the $L_1-C$
filter can capture the original signal in comparison to the HP
filter\footnote{For the simulation of the Ornstein-Uhlenbeck
process, we have chosen $p=0.9985$, $b=20$, $\theta=0.1$ and
$\sigma=2$}.

\subsection{Mixing trend and mean-reverting properties}

We now combine the two schemes proposed above. In this case, we define two
regularization parameters $\lambda _{1}$ and $\lambda _{2}$ corresponding to
two penalty conditions $\sum_{t=1}^{n-1}\left\vert x_{t}-x_{t-1}\right\vert $
and $\sum_{t=2}^{n-1}\left\vert x_{t-1}-2x_{t}+x_{t+1}\right\vert $. Our
objective function for the primal problem becomes now:%
\begin{equation*}
\frac{1}{2}\sum_{t=1}^{n}\left( y_{t}-x_{t}\right) ^{2}+\lambda
_{1}\sum_{t=1}^{n-1}\left\vert x_{t}-x_{t-1}\right\vert +\lambda
_{2}\sum_{t=2}^{n-1}\left\vert x_{t-1}-2x_{t}+x_{t+1}\right\vert
\end{equation*}%
which can be again rewritten in the matrix form:
\begin{equation*}
\frac{1}{2}\left\Vert y-x\right\Vert _{2}^{2}+\lambda _{1}\left\Vert
D_{1}x\right\Vert _{1}+\lambda _{2}\left\Vert D_{2}x\right\Vert _{1}
\end{equation*}%
where the $D_{1}$ and $D_{2}$ operators are respectively the $\left(
n-1\right) \times n$ and $\left( n-2\right) \times n$ matrices defined in
equations (\ref{eq:D1}) and (\ref{eq:D2}).
\bigskip

\begin{figure}[tbph]
\centering
\caption{$L_1-TC$ filtering versus HP filtering for the model (\ref{eq:model1})}
\label{fig:test-model1-tc}
\figureskip
\includegraphics[width = \figurewidth, height = \figureheight]{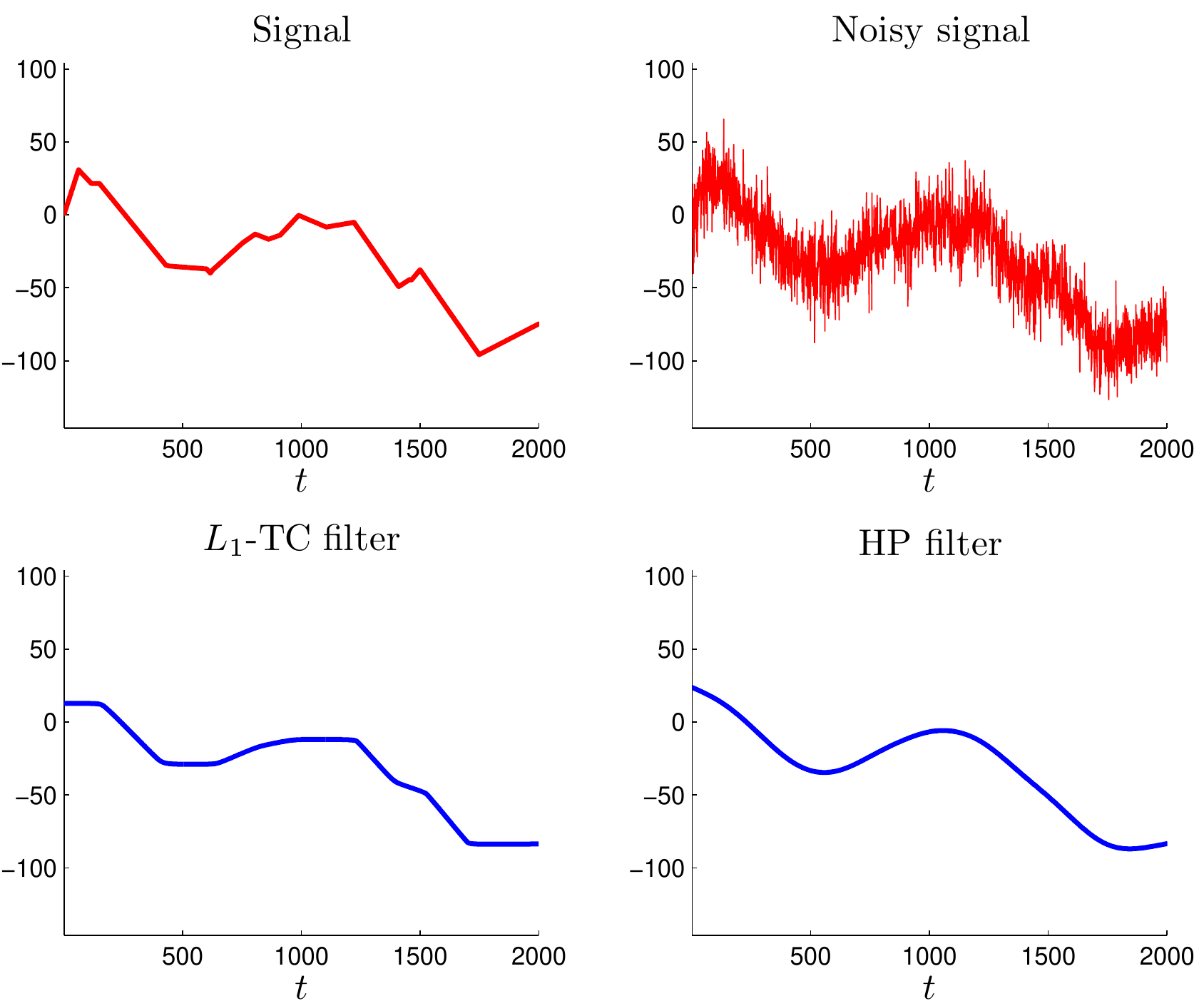}
\end{figure}

\begin{figure}[tbph]
\centering
\caption{$L_1-TC$ filtering versus HP filtering for the model (\ref{eq:model2})}
\label{fig:test-model2-tc}
\figureskip
\includegraphics[width = \figurewidth, height = \figureheight]{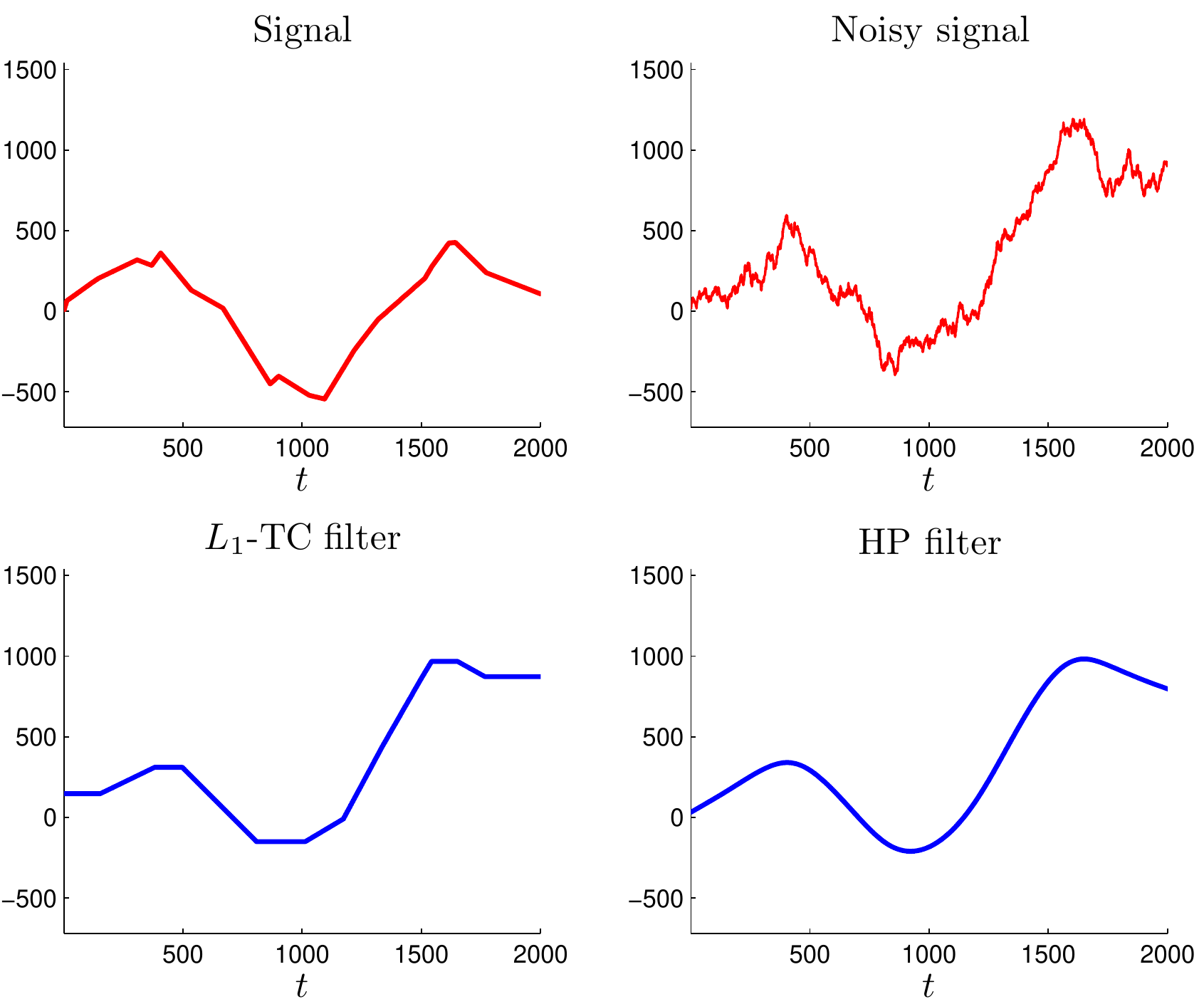}
\end{figure}

In Figures \ref{fig:test-model1-tc} and \ref{fig:test-model2-tc},
we test the efficiency of the mixing scheme on the straight trend lines model
(\ref{eq:model1}) and the random walk model (\ref{eq:model2})%
\footnote{For both models, the parameters are $p=0.99$, $b=0.5$ and
$\sigma=5$.}.


\subsection{How to calibrate the regularization parameters?}\label{subsec:Calibration}

As shown above, the trend obtained from $L_1$ filtering depends on
the parameter $\lambda$ of the regularization procedure. For large
values of $\lambda$, we obtain the long-term trend of the data while
for small values of $\lambda$, we obtain short-term trends of the
data. In this paragraph, we attempt to define a procedure which
permits to do the right choice on the smoothing parameter according
to our need of trend extraction.

\subsubsection{A preliminary remark}

\begin{figure}[tbph]
\centering
\caption{Influence of the smoothing parameter $\lambda$}
\label{fig:lambda-S&P500-1}
\figureskip
\includegraphics[width = \figurewidth, height = \figureheight]{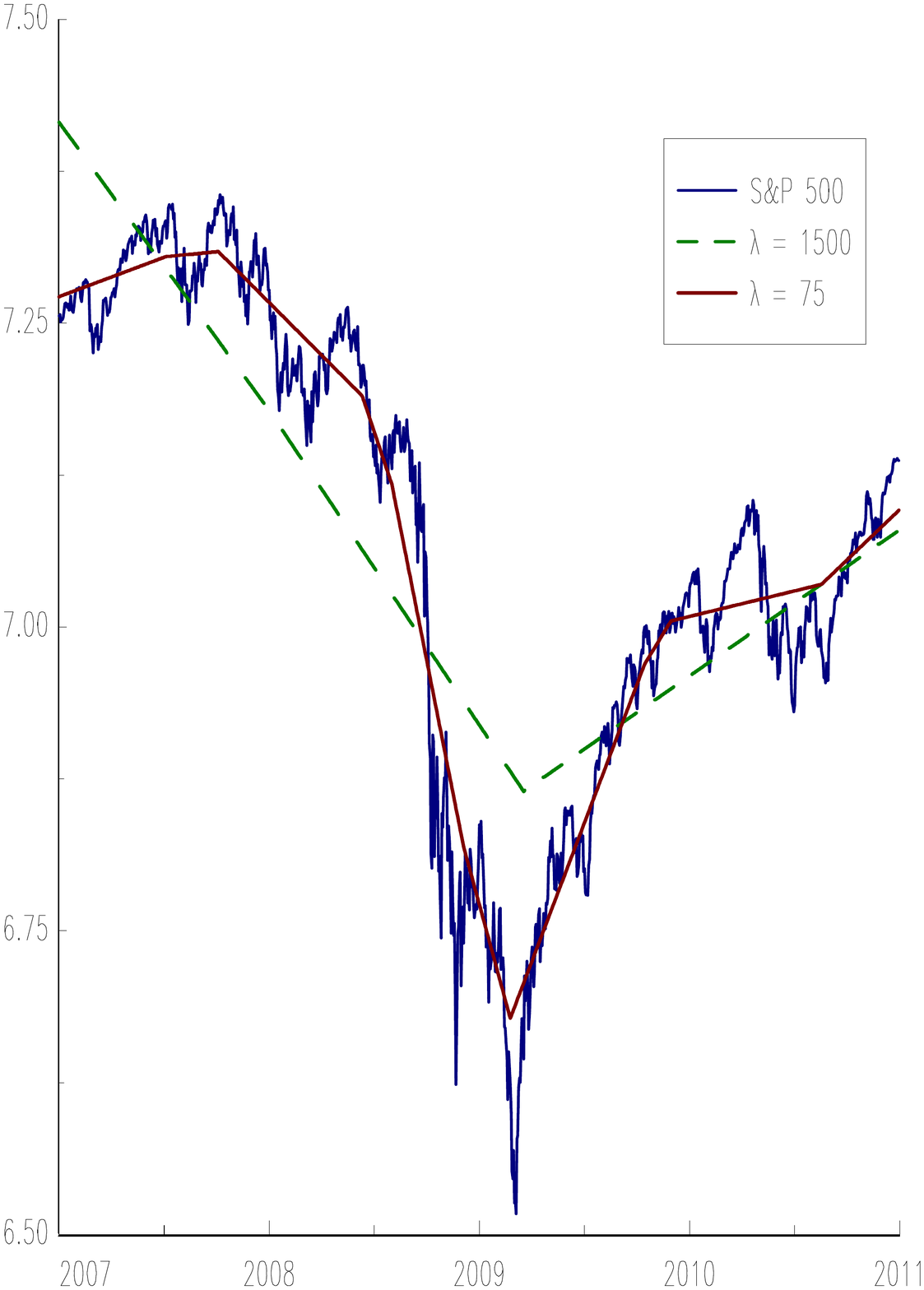}
\end{figure}

For small value of $\lambda$, we recover the original form of the
signal. For large value of $\lambda$, we remark that there exists a maximum
value $\lambda_\textrm{max}$ above which the trend signal has the
affine form:
\begin{equation*}
x_t = \alpha + \beta t
\end{equation*}
where $\alpha$ and $\beta$ are two constants which do not depend on
the time $t$. The value of $\lambda_\textrm{max}$ is given by:
\begin{equation*}
\lambda_\textrm{max} =
\left\|\left(DD^\top\right)^{-1}Dy\right\|_\infty
\end{equation*}
We can use this remark to get an idea about the order of magnitude
of $\lambda$ which should be used to determine the trend over a
certain time period $T$. In order to show this idea, we take the
data over the total period $T$. If we want to have the global trend
on this period, we fix $\lambda=\lambda_\textrm{max}$. This
$\lambda$ will gives the unique trend for the signal over the whole
period. If one need to get more detail on the trend over shorter
periods, we can divide the signal into $p$ time intervals and then
estimate $\lambda$ via the mean value of all the
$\lambda^i_\textrm{max}$ parameter:
\begin{equation*}
\lambda = \frac{1}{p}\sum_{i=1}^{p} \lambda^i_\textrm{max}
\end{equation*}
In Figure \ref{fig:lambda-S&P500-1}, we show the results obtained with $p=2$ ($\lambda = 1\,500$)
and $p=6$ ($\lambda = 75$) on the S\&P 500 index.
\bigskip

\begin{figure}[tbph]
\centering
\caption{Scaling power law of the smoothing parameter $\lambda_{\max}$}
\label{fig:lambda-power-law}
\figureskip
\includegraphics[width = \figurewidth, height = \figureheight]{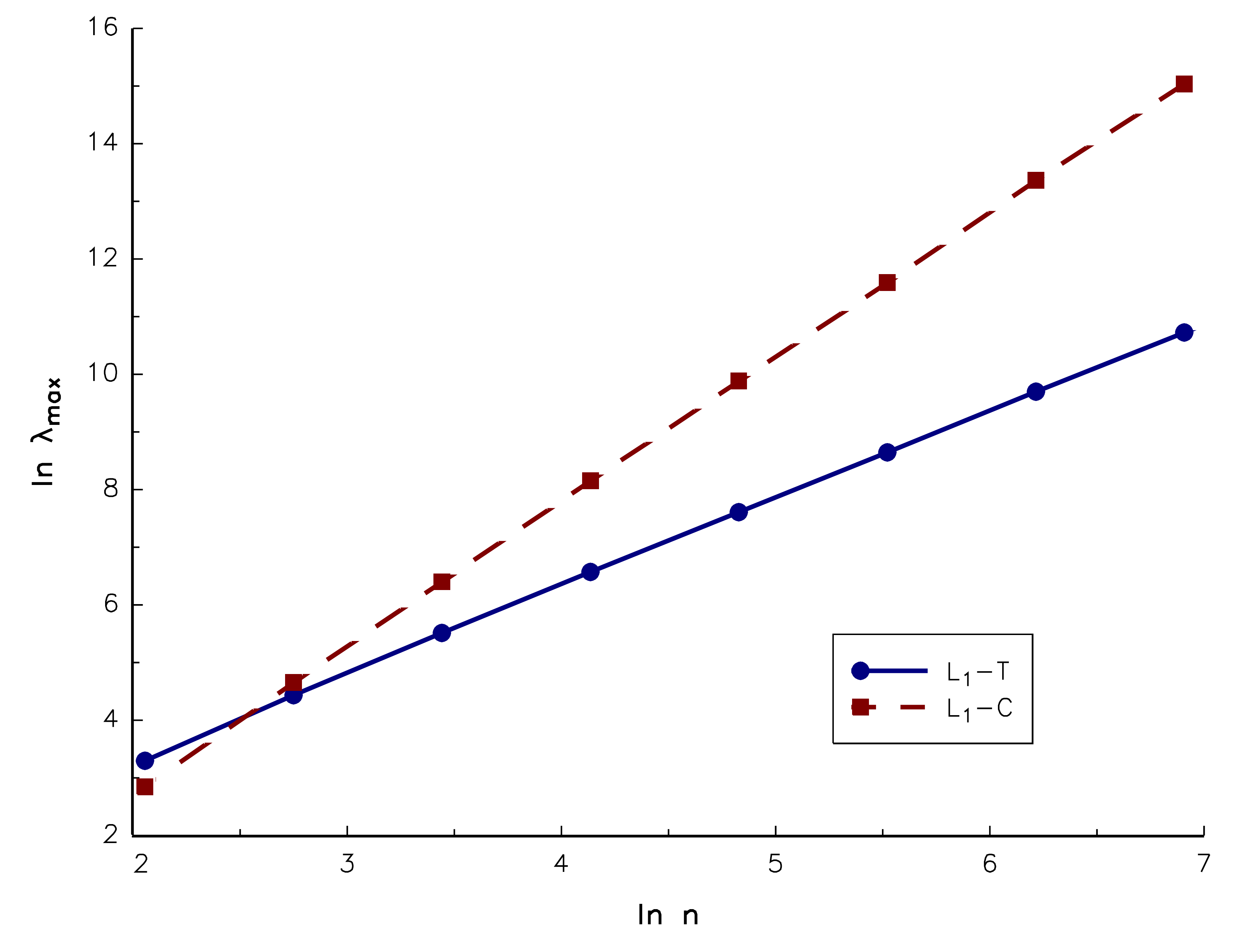}
\end{figure}

Moreover, the explicit calculation of a Brownian motion process
gives us the scaling law of the the smoothing parameter
$\lambda_{\max}$. For the trend filtering scheme, $\lambda_{\max}$
scales as $T^{5/2}$ while for the mean-reverting scheme,
$\lambda_{\max}$ scales as $T^{3/2}$ (see Figure
\ref{fig:lambda-power-law}). Numerical calculation of these powers
for $500$ simulations of the model (\ref{eq:model2}) gives very good
agreement with the analytical result for Brownian motion. Indeed, we
obtain empirically that the power for $L_1-T$ filter is $2.51$
while the one for $L_1-C$ filter is $1.52$.

\subsubsection{Cross validation procedure}

\begin{figure}[h!]
\centering
\caption{Calibration procedure with the S\&P 500 index}
\figureskip
\label{fig:lambda3}
\includegraphics[width = \figurewidth, height = \figureheight]{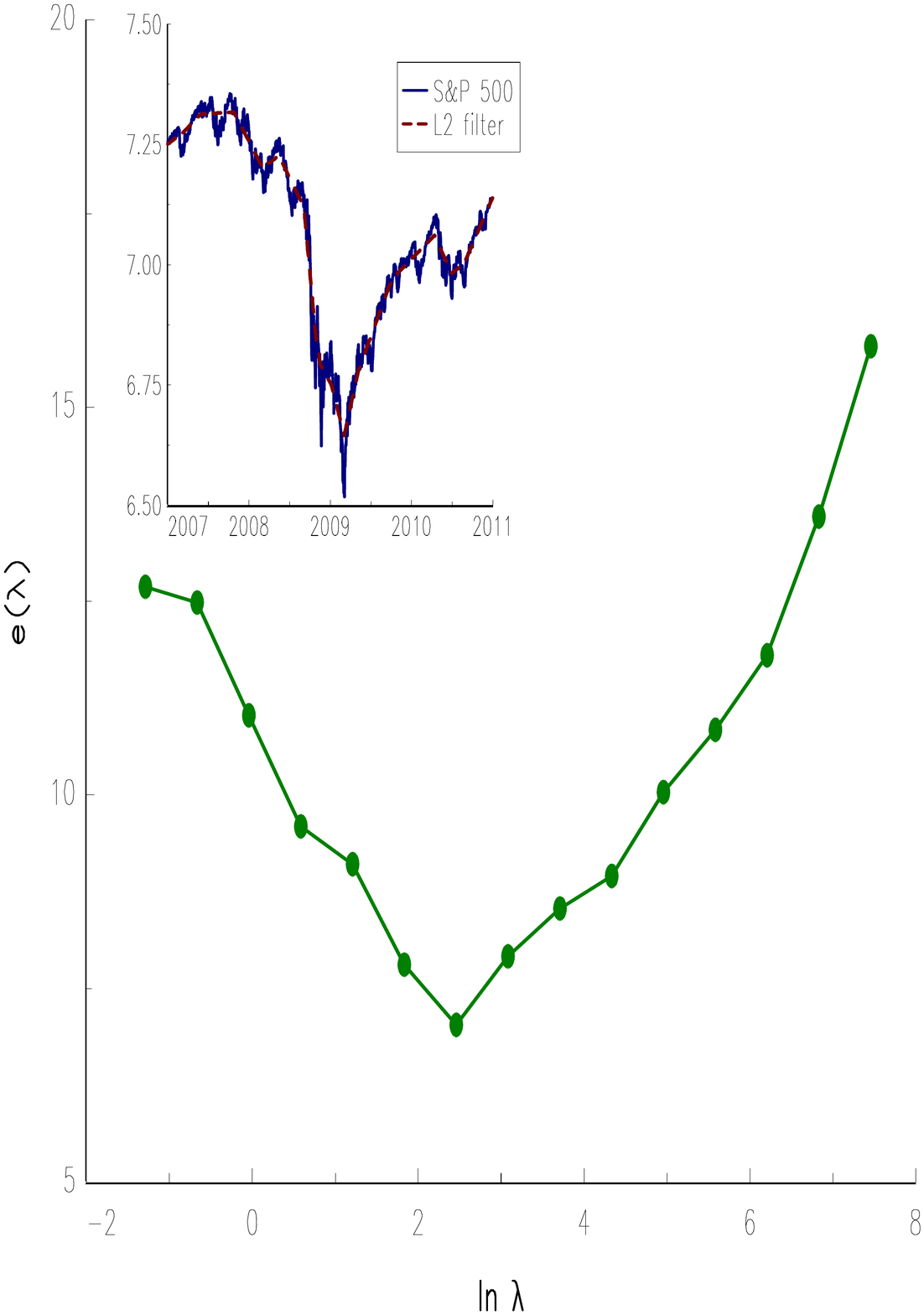}
\end{figure}

In this paragraph, we discuss how to employ a cross-validation
scheme in order to calibrate the smoothing parameter $\lambda$ of
our model. We define two additional parameters which characterize
the trend detection mechanism. The first parameter $T_1$ is the
width of the data windows to estimate the optimal $\lambda$ with
respect to our target strategy. This parameter controls the
precision of our calibration. The second parameter $T_2$ is used to
estimate the prediction error of the trends obtained in the main
window. This parameter characterizes the time horizon of the
investment strategy.
\begin{figure}[tbph]
\centering
\caption{Cross-validation procedure for determining optimal value $\lambda^\star$}
\label{fig:algorithm1}
\figureskip
\begin{center}
\begin{minipage}[t]{0.75\textwidth}
\begin{picture}(400,60)

\put(0,12){$\mid$}
\put(1,15){\vector(1,0){290}}

\put(200,12){$\parallel$}

\put(44,27){$\mid$}
\put(45,30){\vector(1,0){75}}
\put(83,19){\color{RoyalBlue}$T_1$}
\put(47,45){\fcolorbox{RoyalBlue}{White}{\makebox(55,10){\textcolor{RoyalBlue}{Training set}}}}

\put(121,27){$\mid$}
\put(122,30){\vector(1,0){50}}
\put(145,19){\color{Purple}$T_2$}
\put(124,45){\fcolorbox{Purple}{White}{\makebox(35,10){\textcolor{Purple}{Test set}}}}

\put(204,27){$\mid$}
\put(205,30){\vector(1,0){50}}
\put(228,19){\color{ForestGreen}$T_2$}
\put(207,45){\fcolorbox{ForestGreen}{White}{\makebox(50,10){\textcolor{ForestGreen}{Forecasting}}}}

\put(50,0){\textcolor{pink}{Historical data}}
\put(190,0){\textcolor{pink}{Today}}
\put(250,0){\textcolor{pink}{Prediction}}
\end{picture}
\end{minipage}
\end{center}
\end{figure}
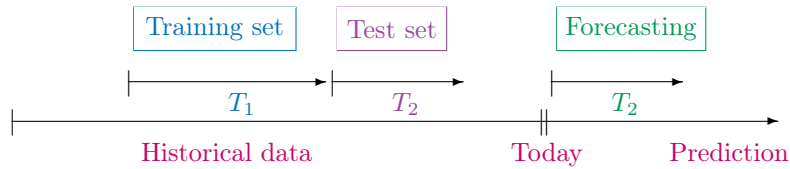
Figure \ref{fig:algorithm1} shows how the data set is divided into
different windows in the cross validation procedure. In order to get
the optimal parameter $\lambda$, we compute the total error after
scanning the whole data by the window $T_1$. The algorithm of this
calibration process is described as following:
\clearpage

\begin{algorithm}[h!]
\caption{Cross validation procedure for $L_1$ filtering}
\label{alg-CV}
\begin{algorithmic}[0]
\Procedure {CV\_Filter$\left(T_1,T_2\right)$}{}
    \State Divide the historical data by $m$ rolling test sets $T_2^i$ ($i=1,\ldots,m$)
    \State For each test window $T_2^i$, compute the statistic $\lambda_{\max}^i$
    \State From the array of $\left( \lambda_{\max}^i \right)$, compute the average $\bar{\lambda}$ and the standard deviation $\sigma_\lambda$
    \State Compute the boundaries $\lambda_1=\bar{\lambda}-2\sigma_\lambda$ and $\lambda_2=\bar{\lambda}+2\sigma_\lambda$
    \For{$j=1:n$}
        \State Compute $\lambda_j = \lambda_{1}\left(\lambda_2/\lambda_1\right)^{\left(j/n\right)}$
        \State Divide the historical data by $p$ rolling training sets $T_1^k$ ($k=1,\ldots,p$)
        \For{$k=1:p$}
          \State For each training window $T_1^k$, run the $L_1$ filter
          \State Forecast the trend for the adjacent test window $T_2^k$
          \State Compute the error $e^k\left(\lambda_j\right)$ on the test window $T_2^k$
        \EndFor
        \State Compute the total error $e\left(\lambda_j\right) = \sum_{k=1}^{m}e^k\left(\lambda_j\right)$
    \EndFor
    \State Minimize the total error $e\left(\lambda\right)$ to find the optimal value $\lambda^\star$
    \State Run the $L_1$ filter with $\lambda = \lambda^\star$
\EndProcedure
\end{algorithmic}
\end{algorithm}
Figure \ref{fig:lambda3} illustrates the calibration procedure for the S\&P 500
index with $T_1 = 400$ and $T_2=50$ for the S\&P 500
index (the number of observations is equal to $1\,008$ trading days). With $m=p=12$ and $n=15$, the estimated
optimal value $\lambda^{\star}$ for the $L_1-T$ filter is equal to $7.03$.

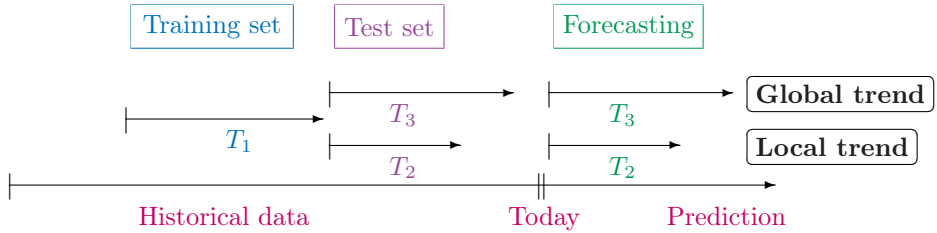
\begin{figure}[hb]
\centering
\caption{Cross validation procedure for two-trend model}
\label{fig:algorithm2}
\figureskip
\begin{center}
\begin{minipage}[t]{0.85\textwidth}

\begin{picture}(400,90)

\put(0,12){$\mid$}
\put(1,15){\vector(1,0){290}}

\put(200,12){$\parallel$}

\put(44,37){$\mid$}
\put(45,40){\vector(1,0){75}}
\put(83,29){\color{RoyalBlue}$T_1$}
\put(47,70){\fcolorbox{RoyalBlue}{White}{\makebox(55,10){\textcolor{RoyalBlue}{Training set}}}}

\put(121,27){$\mid$}
\put(122,30){\vector(1,0){50}}
\put(145,19){\color{Purple}$T_2$}
\put(121,47){$\mid$}
\put(122,50){\vector(1,0){70}}
\put(145,39){\color{Purple}$T_3$}
\put(124,70){\fcolorbox{Purple}{White}{\makebox(35,10){\textcolor{Purple}{Test set}}}}

\put(204,27){$\mid$}
\put(205,30){\vector(1,0){50}}
\put(228,19){\color{ForestGreen}$T_2$}
\put(204,47){$\mid$}
\put(205,50){\vector(1,0){70}}
\put(228,39){\color{ForestGreen}$T_3$}
\put(207,70){\fcolorbox{ForestGreen}{White}{\makebox(50,10){\textcolor{ForestGreen}{Forecasting}}}}

\put(50,0){\textcolor{pink}{Historical data}}
\put(190,0){\textcolor{pink}{Today}}
\put(250,0){\textcolor{pink}{Prediction}}

\put(280,26){\ovalbox{\textcolor{Black}{\textbf{Local trend}}}}
\put(280,46){\ovalbox{\textcolor{Black}{\textbf{Global trend}}}}

\end{picture}
\end{minipage}
\end{center}
\end{figure}

We have observed that this calibration procedure is more favorable
for long-term time horizon, that is to estimate a global trend. For
short-term time horizon, the prediction of local trends is much more
perturbed by the noise. We have computed the probability of having
good prediction on the tendency of the market for long-term and
short-term time horizons. This probability is about $70\%$ for 3
months time horizon while it is just $50\%$ for one week time
horizon. It comes that even if the fit is good for the past, the
noise is however large meaning that the prediction of the future
tendency is just $\nicefrac{1}{2}$ for an increasing market and
$\nicefrac{1}{2}$ for a decreasing market. In order to obtain better
results for smaller time horizons, we improve the last algorithm by
proposing a two-trend model. The first trend is the local one which
is determined by the first algorithm with the parameter $T_2$
corresponding to the local prediction. The second trend is the
global one which gives the tendency of the market over a longer
period $T_3$. The choice of this global trend parameter is very
similar to the choice of the moving-average parameter. This model
can be considered as a simple version of mean-reverting model for
the trend. In Figure \ref{fig:algorithm2}, we describe how the data
set is divided for estimating the local trend and the global trend.

The procedure for estimating the trend of the signal in the
two-trend model is summarized in Algorithm \ref{alg-FP}. The
corrected trend is now determined by studying the relative position
of the historical data to the global trend. The reference
position is characterized by the standard deviation
$\sigma\left(y_t-x^G_t\right)$ where $x^G_t$ is the filtered global trend.

\begin{algorithm}[ht]
\caption{Prediction procedure for the two-trend model}
\label{alg-FP}
\begin{algorithmic}[0]
\Procedure {Predict\_Filter$\left(T_l,T_g\right)$}{}
    \State Compute the local trend $x_t^{L}$ for the time horizon $T_2$ with the
    CV\_FILTER procedure
    \State Compute the global trend $x_t^G$ for the time horizon $T_3$ with the
    CV\_FILTER procedure
    \State Compute the standard deviation $\sigma\left(y_t-x_t^G\right)$ of data with respect to the global trend
    \If{$\left|y_t - x_t^G\right| < \sigma\left(y_t-x_t^G\right)$}
            \State Prediction $\leftarrow$ $x_t^L$
    \Else
            \State Prediction $\leftarrow$ $x_t^G$
        \EndIf
\EndProcedure
\end{algorithmic}
\end{algorithm}

\section{Application to momentum strategies}

In this section, we apply the previous framework to the S\&P 500
index. First, we illustrate the calibration procedure for a given
trading date. Then, we backtest a momentum strategy by estimating
dynamically the optimal filters.

\subsection{Estimating the optimal filter for a given trading date}

\begin{figure}[tbph]
\centering
\caption{Comparison between different $L_1$ filters on S\&P 500 Index}
\figureskip
\label{fig:backtest1}
\includegraphics[width = \figurewidth, height = \figureheight]{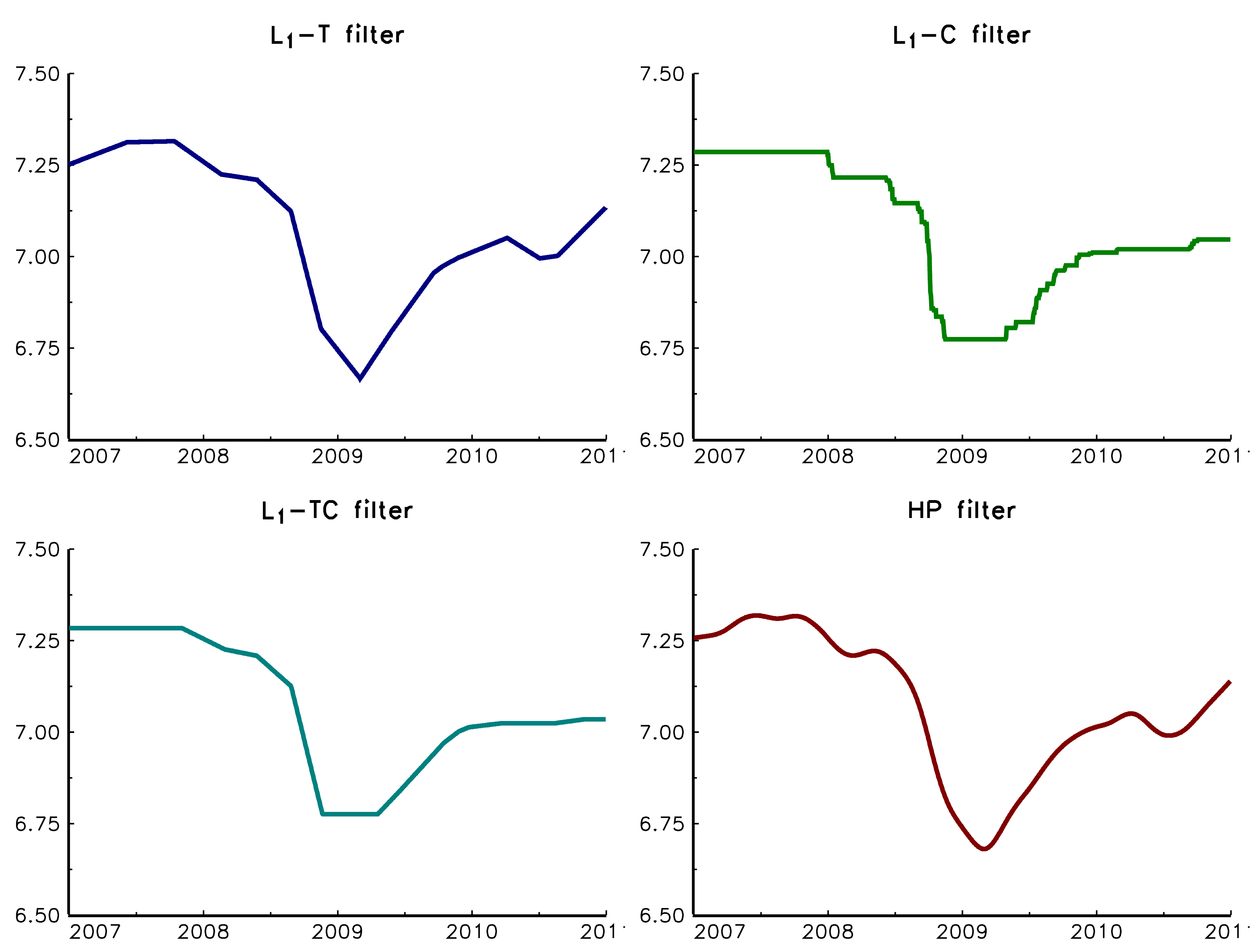}
\end{figure}

We would like to estimate the optimal filter for January 3rd, 2011
by considering the period from January 2007 to December 2010. We use
the previous algorithms with $T_1 = 400$ and $T_2 = 50$. The optimal
parameters are $\lambda_1 = 2.46$ (for the $L_1-C$ filter) and
$\lambda_2 = 15.94$ (for the $L_2-T$ filter). Results are reported
in Figure \ref{fig:backtest1}. The trend for the next 50 trading
days is estimated to $7.34\%$ for the $L_1-T$ filter and $7.84\%$
for the HP filter whereas it is null for the $L_1-C$ and $L_1-TC$
filters. By comparison, the true performance of the S\&P 500 index
is $1.90\%$ from January 3rd, 2011 to March 15th, 2011%
\footnote{It corresponds exactly to a period of 50 trading days}.

\subsection{Backtest of a momentum strategy}

\subsubsection{Design of the strategy}

Let us consider a class of self-financed strategies on a risky asset
$S_{t}$ and a risk-free asset $B_{t}$. We assume that the dynamics
of these assets is:
\begin{eqnarray*}
\mathrm{d}B_t   & = & r_t B_t\, \mathrm{d}t \\
\mathrm{d}S_{t} & = & \mu_t S_t \, \mathrm{d}t + \sigma_t S_t\, \mathrm{d}W_t
\end{eqnarray*}
where $r_t$ is the risk-free rate, $\mu_t$ is the trend of the asset price and $\sigma_t$ is the volatility.
We denote $\alpha _{t}$ the proportion of investment in the risky
asset and $\left( 1-\alpha _{t}\right)$ the part invested in the risk-free asset. We start with an
initial budget $W_{0}$ and expect a final wealth $W_{T}$. The optimal
strategy is the one which optimizes the expectation of the utility function $%
U\left( W_{T}\right) $ which is increasing and concave. It is equivalent to
the Markowitz problem which consists of maximizing the wealth of the portfolio under a
penalty of risk:%
\begin{equation*}
\sup_{\alpha \in \mathbb{R}}\left\{ \mathbb{E}\left( W_{T}^{\alpha }\right) -%
\frac{\lambda }{2}\sigma ^{2}\left( W_{T}^{\alpha }\right) \right\}
\end{equation*}%
which is equivalent to:%
\begin{equation*}
\sup_{\alpha \in \mathbb{R}}\left\{ \alpha _{t}\mu _{t}-\frac{\lambda }{2}%
W_{0}\alpha _{t}^{2}\sigma _{t}^{2}\right\}
\end{equation*}%
As the objective function is concave, the maximum corresponds to the zero
point of the gradient $\mu _{t}-\lambda W_{0}\alpha _{t}\sigma _{t}^{2}$. We
obtain the optimal solution:%
\begin{equation*}
\alpha _{t}^{\star }=\frac{1}{\lambda W_{0}}\frac{\mu _{t}}{\sigma _{t}^{2}}
\end{equation*}%
In order to limit the explosion of $\alpha _{t}$, we also impose the
following constraint $\alpha _{\min }\leq \alpha _{t}\leq \alpha _{\max }$:
\begin{equation*}
\alpha _{t}^{\star }=\max \left( \min \left( \frac{1}{\lambda W_{0}}\frac{%
\mu _{t}}{\sigma _{t}^{2}},\alpha _{\min }\right) ,\alpha _{\max }\right)
\end{equation*}%
The wealth of the portfolio is then given by the following expression:
\begin{equation*}
W_{t+1}=W_{t}+W_{t}\left( \alpha _{t}^{\star }\left( \frac{S_{t+1}}{S_{t}}%
-1\right) +(1-\alpha_{t}^{\star })r_{t}\right)
\end{equation*}

\subsubsection{Results}

In the following simulations, we use the estimators $\hat{\mu}_{t}$ and $%
\hat{\sigma}_{t}$ in place of $\mu _{t}$ and $\sigma _{t}$. For $\hat{\mu}%
_{t}$, we consider different models like $L_{1}$, HP and
moving-average filters\footnote{%
We note them respectively $\hat{\mu}_{t}^{L_{1}}$, $\hat{\mu}_{t}^{\mathrm{HP%
}}$ and $\hat{\mu}_{t}^{\mathrm{MA}}$.} whereas we use the following
estimator for the volatility:
\begin{equation*}
\hat{\sigma}_{t}^{2}=\frac{1}{T}\int_{0}^{T}\sigma _{t}^{2}\,\mathrm{d}t=%
\frac{1}{T}\sum_{i=t-T+1}^{t}\ln ^{2}\frac{S_{i}}{S_{i-1}}
\end{equation*}%
We consider a long/short strategy, that is $\left( \alpha _{\min },\alpha
_{\max }\right) =\left( -1,1\right) $. In the particular case of the $\hat{%
\mu}_{t}^{L_{1}}$ estimator, we consider three different models:

\begin{enumerate}
\item the first one is based on the local trend;

\item the second one is based on the global trend;

\item the combination of both local and global trends corresponds to the
third model.
\end{enumerate}

\noindent For all these strategies, the test set of the local trend $T_{2}$ is equal to 6
months (or 130 trading days) whereas the length of the test set for global trend is four
times the length of the test set -- $T_{3}=4T_{2}$ -- meaning that $T_{3}$
is one year (or 520 trading days). This choice of $T_3$ agrees with the habitual choice of the
width of the windows in moving average estimator. The length of the training set is also four
times the length of the test set $T_1$. The study period is from January 1998
to December 2010. In the backtest, the trend estimation is updated every day.
In Table \ref{tab:backtest1}, we summarize the results obtained with the
different models cited above for the backtest. We remark that the best performances correspond to the
case of global trend, HP and two-trend models. Because HP filter is calibrated to the window of the moving-average
filter which is equal to $T_3$, it is  not surprising that
the performances of these three models are similar. On the considered period of the backtest, the S\&P
does not have a clear upward or downward trend. Hence, the local trend estimator does not give a good prediction
and this strategy gives the worst performance. By contrast, the two-trend model takes into account the trade-off
between local trend and global trend and gives a better result
\begin{table}
\centering
\caption{Results for the Backtest}
\tableskip
\label{tab:backtest1}
\begin{tabular}{cc|ccccc}
Model                         & Trend & Performance & Volatility & Sharpe      & IR     & Drawdown     \\ \hline
S\&P 500                      &       & $2.04\%$    & $21.83\%$  & $    -0.06$ &        & $56.78$      \\
$\hat{\mu}_{t}^{\mathrm{MA}}$ &       & $3.13\%$    & $18.27\%$  & $    -0.01$ & $0.03$ & $33.83$      \\
$\hat{\mu}_{t}^{\mathrm{HP}}$ &       & $6.39\%$    & $18.28\%$  & ${\bN}0.17$ & $0.13$ & $39.60$      \\
$\hat{\mu}_{t}^{L_{1}}$       & (LT)  & $3.17\%$    & $17.55\%$  & $    -0.01$ & $0.03$ & $25.11$      \\
$\hat{\mu}_{t}^{L_{1}}$       & (GT)  & $6.95\%$    & $19.01\%$  & ${\bN}0.19$ & $0.14$ & $31.02$      \\
$\hat{\mu}_{t}^{L_{1}}$       & (LGT) & $6.47\%$    & $18.18\%$  & ${\bN}0.17$ & $0.13$ & $31.99$
\end{tabular}
\end{table}

\section{Extension to the multivariate case}

We now extend the $L_{1}$ filtering scheme to a multivariate time series $%
y_{t}=\left( y_{t}^{(1)},\ldots ,y_{t}^{(m)}\right) $. The underlying idea
is to estimate the common trend of several univariate time series. In
finance, the time series correspond to the prices of several assets.
Therefore, we can build long/short strategies between these assets by
comparing the individual trends and the common trend.\bigskip

For the sake of simplicity, we assume that all the signals are rescaled to
the same order of magnitude\footnote{%
For example, we may center and standardize the time series by subtracting
the mean and dividing by the standard deviation.}. The objective function
becomes new:%
\begin{equation*}
\frac{1}{2}\sum_{i=1}^{m}\left\Vert y^{(i)}-x\right\Vert^2 _{2}+\lambda
\left\Vert Dx\right\Vert _{1}
\end{equation*}%
In Appendix \ref{appendix-dual-L1-TM}, we show that this problem is
equivalent to the $L_{1}$ univariate problem by considering $\bar{y}%
_{t}=m^{-1}\sum_{i=1}^{m}y^{\left( i\right) }$ as the signal.

\section{Conclusion}

Momentum strategies are efficient ways to use the market tendency for
building trading strategies. Hence, a good estimator of the trend is
essential from this perspective. In this paper, we show that we can use $%
L_{1}$ filters to forecast the trend of the market in a very simple way. We
also propose a cross-validation procedure to calibrate the optimal
regularization parameter $\lambda $ where the only information to provide is
the investment time horizon. More sophisticated models based on a local and
global trends is also discussed. We remark that these models can reflect the
effect of mean-reverting to the global trend of the market. Finally, we
consider several backtests on the S\&P 500 index and obtain competing
results with respect to the traditional moving-average filter.

\clearpage

\appendix

\section{Computational aspects of $L_1$, $L_2$ filters}

\subsection{The dual problem\label{appendix-dual-problem}}

\subsubsection{The $L_1-T$ filter\label{appendix-dual-L1-T}}

This problem can be solved by considering the dual problem which is a QP
program. We first rewrite the primal problem with new variable $z=Dx$:%
\begin{eqnarray*}
&\min &\frac{1}{2}\left\Vert y-x\right\Vert _{2}^{2}+\lambda \left\Vert
z\right\Vert _{1} \\
&\text{u.c.}&z=Dx
\end{eqnarray*}%
We construct now the Lagrangian function with the dual variable $\nu \in
\mathbb{R}^{n-2}$:%
\begin{equation*}
\mathcal{L}\left( x,z,\nu \right) =\frac{1}{2}\left\Vert y-x\right\Vert
_{2}^{2}+\lambda \left\Vert z\right\Vert _{1}+\nu ^{\top }\left( Dx-z\right)
\end{equation*}%
The dual objective function is obtained in the following way:%
\begin{equation*}
\inf\nolimits_{x,z}\mathcal{L}\left( x,z,\nu \right) =-\frac{1}{2}\nu ^{\top
}DD^{\top }\nu +y^{\top }D^{\top }\nu
\end{equation*}%
for $-\lambda \mathbf{1}\leq \nu \leq \lambda \mathbf{1}$. According to the
Kuhn-Tucker theorem, the initial problem is equivalent to the dual problem:%
\begin{eqnarray*}
&\min &\frac{1}{2}\nu ^{\top }DD^{\top }\nu -y^{\top }D^{\top }\nu  \\
&\text{u.c.}&-\lambda \mathbf{1}\leq \nu \leq \lambda \mathbf{1}
\end{eqnarray*}%
This QP program can be solved by traditional Newton algorithm or by
interior-point methods, and the final solution of the trend reads
\begin{equation*}
x^{\star }=y-D^{\top }\nu
\end{equation*}

\subsubsection{The $L_1-C$ filter\label{appendix-dual-L1-C}}

The optimization procedure for $L_{1}-C$ filter follows the same strategy as
the $L_{1}-T$ filter. We obtain the same quadratic program with the $D$
operator replaced by $\left( n-1\right) \times n$ matrix which is the
discrete version of the first order derivative:%
\begin{equation*}
D=\left[
\begin{array}{rrrrrr}
-1 & 1 & 0 &  &  &  \\
0 & -1 & 1 & 0 &  &  \\
&  &  & \ddots  &  &  \\
&  &  & -1 & 1 & 0 \\
&  &  &  & -1 & 1%
\end{array}%
\right]
\end{equation*}

\subsubsection{The $L_1-TC$ filter\label{appendix-dual-L1-TC}}

In order to follow the same strategy presented above, we introduce two
additional variables $z_{1}=D_{1}x$ and $z_{2}=D_{2}x$. The initial problem
becomes:%
\begin{eqnarray*}
&\min &\frac{1}{2}\left\Vert y-x\right\Vert _{2}^{2}+\lambda _{1}\left\Vert
z_{1}\right\Vert _{1}+\lambda _{2}\left\Vert z_{2}\right\Vert _{1} \\
&\text{u.c.}&\left\{
\begin{array}{c}
z_{1}=D_{1}x \\
z_{2}=D_{2}x%
\end{array}%
\right.
\end{eqnarray*}%
The Lagrangian function with the dual variables $\nu _{1}\in \mathbb{R}^{n-1}
$ and $\nu _{2}\in \mathbb{R}^{n-2}$\ is:%
\begin{equation*}
\mathcal{L}\left( x,z_{1},z_{2},\nu _{1},\nu _{2}\right) =\frac{1}{2}%
\left\Vert y-x\right\Vert _{2}^{2}+\lambda _{1}\left\Vert z_{1}\right\Vert
_{1}+\lambda _{2}\left\Vert z_{2}\right\Vert _{1}+\nu _{1}^{\top }\left(
D_{1}x-z_{1}\right) +\nu _{2}^{\top }\left( D_{2}x-z_{2}\right)
\end{equation*}%
whereas the dual objective function is:%
\begin{equation*}
\inf\nolimits_{x,z_{1},z_{2}}\mathcal{L}\left( x,z_{1},z_{2},\nu _{1},\nu
_{2}\right) =-\frac{1}{2}\left\Vert D_{1}^{\top }\nu _{1}+D_{2}^{\top }\nu
_{2}\right\Vert _{2}^{2}+y^{\top }\left( D_{1}^{\top }\nu _{1}+D_{2}^{\top
}\nu _{2}\right)
\end{equation*}%
for $-\lambda _{i}\mathbf{1}\leq \nu _{i}\leq \lambda _{i}\mathbf{1}$ ($i=1,2
$). Introducing the variable $z=\left( z_{1},z_{2}\right) $ and $\nu =\left(
\nu _{1},\nu _{2}\right) $, the initial problem is equivalent to the dual
problem:%
\begin{eqnarray*}
&\min &\frac{1}{2}\nu ^{\top }Q\nu -R^{\top }\nu  \\
&\text{u.c.}&-\nu ^{+}\leq \nu \leq \nu ^{+}
\end{eqnarray*}%
with $D=\left(
\begin{array}{c}
D_{1} \\
D_{2}%
\end{array}%
\right) $, $Q=DD^{\top }$, $R=Dy$ and $\nu ^{+}=\left(
\begin{array}{c}
\lambda _{1} \\
\lambda _{2}%
\end{array}%
\right) \mathbf{1}$. The solution of the primal problem is then given by $%
x^{\star }=y-D^{\top }\nu $.

\subsubsection{The $L_{1}-T$ multivariate filter\label{appendix-dual-L1-TM}}

As in the univariate case, this problem can be solved by considering the
dual problem which is a QP program. The primal problem is:%
\begin{eqnarray*}
& \min & \frac{1}{2}\sum_{i=1}^{m}\left\Vert y^{\left( i\right) }-x\right\Vert^2
_{2}+\lambda \left\Vert z\right\Vert _{1} \\
& \text{u.c.} & z=Dx
\end{eqnarray*}%
Let us define $\bar{y}=\left( \bar{y}_{t}\right) $ with $\bar{y}%
_{t}=m^{-1}\sum_{i=1}^{m}y^{\left( i\right) }$. The dual objective function
becomes:%
\begin{equation*}
\inf\nolimits_{x,z}\mathcal{L}\left( x,z,\nu \right) =-\frac{1}{2}\nu ^{\top
}DD^{\top }\nu +\bar{y}^{\top }D^{\top }\nu +\frac{1}{2}\sum_{i=1}^{m}\left(
y^{\left( i\right) }-\bar{y}\right) ^{\top }\left( y^{\left( i\right) }-\bar{%
y}\right)
\end{equation*}%
for $-\lambda \mathbf{1}\leq \nu \leq \lambda \mathbf{1}$. According to the
Kuhn-Tucker theorem, the initial problem is equivalent to the dual problem:%
\begin{eqnarray*}
&\min &\frac{1}{2}\nu ^{\top }DD^{\top }\nu -\bar{y}^{\top }D^{\top }\nu  \\
&\text{u.c.}&-\lambda \mathbf{1}\leq \nu \leq \lambda \mathbf{1}
\end{eqnarray*}%
This QP program can be solved by traditional Newton algorithm or by
interior-point methods and the solution is:
\begin{equation*}
x^{\star }=\bar{y}-D^{\top }\nu
\end{equation*}

\subsection{The interior-point algorithm\label{appendix-interior-point}}

We present briefly the interior-point algorithm of Boyd and Vandenberghe
(2009) in the case of the following optimization problem:%
\begin{eqnarray*}
&\min &f_{0}\left( x\right)  \\
&\text{u.c.}&\left\{
\begin{array}{l}
Ax=b \\
f_{i}\left( x\right) <0\quad \text{for }i=1,\dots ,m%
\end{array}%
\right.
\end{eqnarray*}%
where $f_{0},\dots ,f_{m}:\mathbb{R}^{n}\rightarrow \mathbb{R}$ are convex
and twice continuously differentiable and $\limfunc{rank}\left( A\right) =p<n
$. The inequality constraints will become implicit if one rewrite the
problem as:%
\begin{eqnarray*}
&\min &f_{0}\left( x\right) +\sum_{i=1}^{m}\mathcal{I}_{-}\left( f_{i}\left(
x\right) \right)  \\
&\text{u.c.}&Ax=b
\end{eqnarray*}%
where $\mathcal{I}_{-}\left( u\right) :\mathbb{R}\rightarrow \mathbb{R}$ is
the non-positive indicator function\footnote{%
We have:%
\begin{equation*}
\mathcal{I}_{-}\left( u\right) =\left\{
\begin{array}{ll}
0 & u\leq 0 \\
\infty  & u>0%
\end{array}%
\right.
\end{equation*}%
}. This indicator function is discontinuous, hence the Newton method can not
be applied. In order to overcome this problem, we approximate $\mathcal{I}%
_{-}\left( u\right) $ by the logarithmic barrier function $\mathcal{I}%
_{-}^{\star }\left( u\right) =-\tau ^{-1}\ln \left( -u\right) $ with $\tau
\rightarrow \infty $. Finally the Kuhn-Tucker condition for this
approximation problem gives $r_{t}\left( x,\lambda ,\nu \right) =0$ with:%
\begin{equation*}
r_{\tau }\left( x,\lambda ,\nu \right) =\left(
\begin{array}{c}
\nabla f_{0}\left( x\right) +\nabla f\left( x\right) ^{\top }\lambda
+A^{\top }\nu  \\
-\func{diag}\left( \lambda \right) f\left( x\right) -\tau ^{-1}\mathbf{1} \\
Ax-b%
\end{array}%
\right)
\end{equation*}%
The solution of $r_{\tau }\left( x,\lambda ,\nu \right) =0$ can be obtained
by Newton's iteration for the triple $y=\left( x,\lambda ,\nu \right) $:%
\begin{equation*}
r_{\tau }\left( y+\Delta y\right) \simeq r_{\tau }\left( y\right) +\nabla
r_{\tau }\left( y\right) \Delta y=0
\end{equation*}
This equation gives the Newton's step $\Delta y=-\nabla r_{\tau }\left(
y\right) ^{-1}r_{\tau }\left( y\right) $ which defines the search direction.

\subsection{The scaling of smoothing parameter of $L_1$ filter}

We can try to estimate the order of magnitude of the parameter
$\lambda _{\max }$ by considering the continuous case. Assuming that
the signal is a process $W_{t}$. The value of $\lambda _{\max }$ in
the discrete case
defined by:%
\begin{equation*}
\lambda _{\max }=\left\Vert \left( DD^{\top }\right) ^{-1}Dy\right\Vert
_{\infty }
\end{equation*}%
can be considered as the first primitive $I_{1}\left( T\right)
=\int_{0}^{T}W_{t}\,\mathrm{d}t$ of the process $W_{t}$ if $D=D_{1}$\ ($%
L_{1}-C$ filtering) or the second primitive $I_{2}\left( T\right)
=\int_{0}^{T}\int_{0}^{t}W_{s}\,\mathrm{d}s\,\mathrm{d}t$ of $W_{t}$ if $%
D=D_{2}$\ ($L_{1}-T$ filtering). We have:%
\begin{eqnarray*}
I_{1}\left( T\right)  &=&\int_{0}^{T}W_{t}\,\mathrm{d}t \\
&=&W_{T}T-\int_{0}^{T}t\,\mathrm{d}W_{t} \\
&=&\int_{0}^{T}\left( T-t\right) \,\mathrm{d}W_{t}
\end{eqnarray*}%
The process $I_{1}\left( T\right) $ is a Wiener integral (or a Gaussian
process) with variance:%
\begin{equation*}
\mathbb{E}\left[ I_{1}^{2}\left( T\right) \right] =\int_{0}^{T}\left(
T-t\right) ^{2}\,\mathrm{d}t=\frac{T^{3}}{3}
\end{equation*}%
In this case, we expect that $\lambda _{\max }\sim T^{3/2}$. The second
order primitive can be calculated in the following way:%
\begin{align*}
I_{2}\left( T\right) & =\int_{0}^{T}I_{1}\left( t\right) \,\mathrm{d}t \\
& =I_{1}\left( T\right) T-\int_{0}^{T}t\,\mathrm{d}I_{1}\left( T\right)  \\
& =I_{1}\left( T\right) T-\int_{0}^{T}tW_{t}\,\mathrm{d}t \\
& =I_{1}\left( T\right) T-\frac{T^{2}}{2}W_{T}+\int_{0}^{T}\frac{t^{2}}{2}\,%
\mathrm{d}W_{t} \\
& =-\frac{T^{2}}{2}W_{T}+\int_{0}^{T}\left( T^{2}-Tt+\frac{t^{2}}{2}\right)
\,\mathrm{d}W_{t} \\
& =\frac{1}{2}\int_{0}^{T}\left( T-t\right) ^{2}\,\mathrm{d}W_{T}
\end{align*}%
This quantity is again a Gaussian process with variance:%
\begin{equation*}
\mathbb{E}[I_{2}^{2}\left( T\right) ]=\frac{1}{4}\int_{0}^{T}\left(
T-t\right) ^{4}\,\mathrm{d}t=\frac{T^{5}}{20}
\end{equation*}%
In this case, we expect that $\lambda _{\max }\sim T^{5/2}$.

\subsection{Calibration of the $L_2$ filter}\label{L2Calibration}

\begin{figure}[tbph]
\centering
\caption{Spectral density of moving-average and $L_2$ filters}
\figureskip
\label{fig:lambda4-a}
\includegraphics[width = \figurewidth, height = \figureheight]{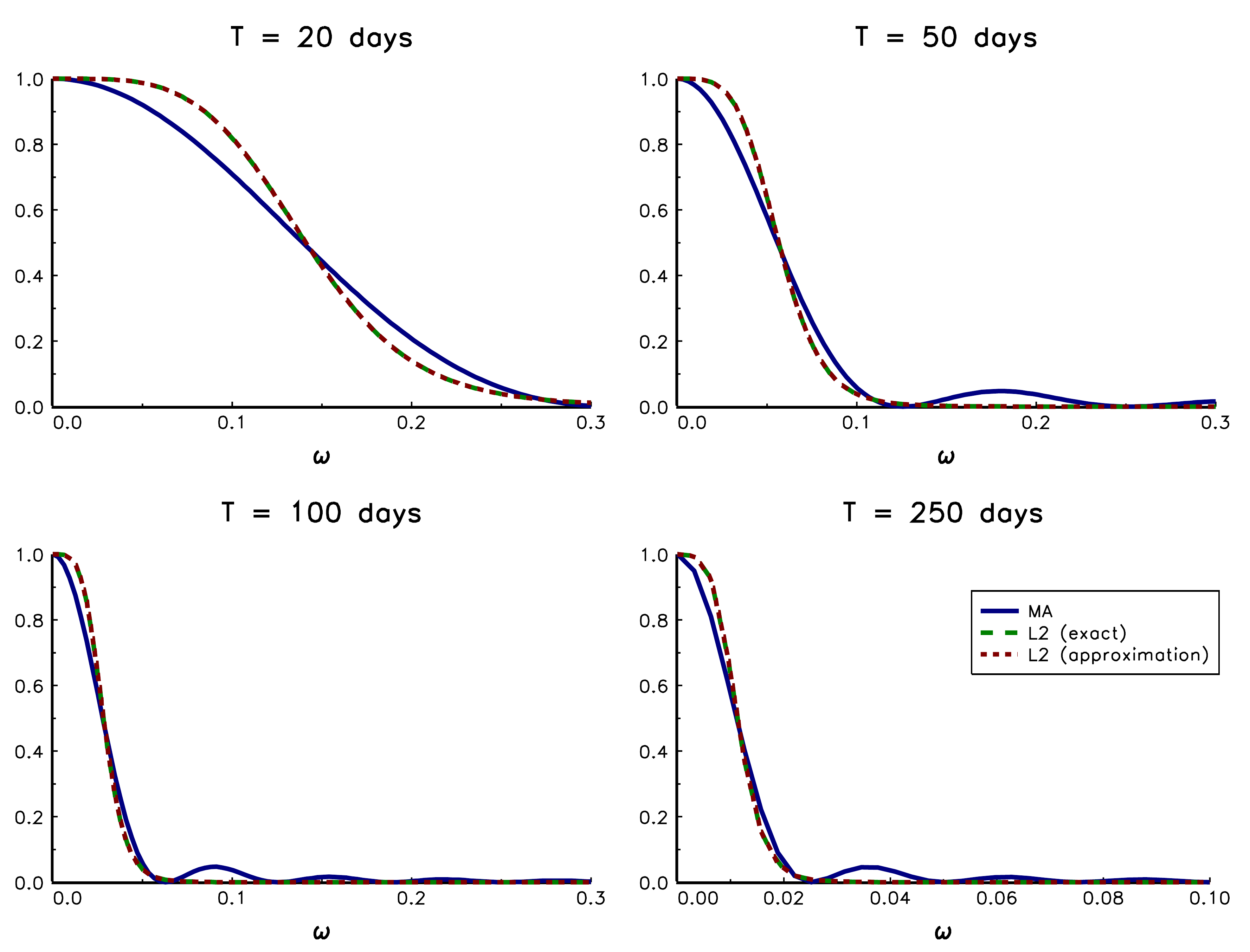}
\figureskip
\centering
\caption{Relationship between the value of $\lambda$ and the length of the moving-average filter}
\figureskip
\label{fig:lambda4-b}
\includegraphics[width = \figurewidth, height = \figureheight]{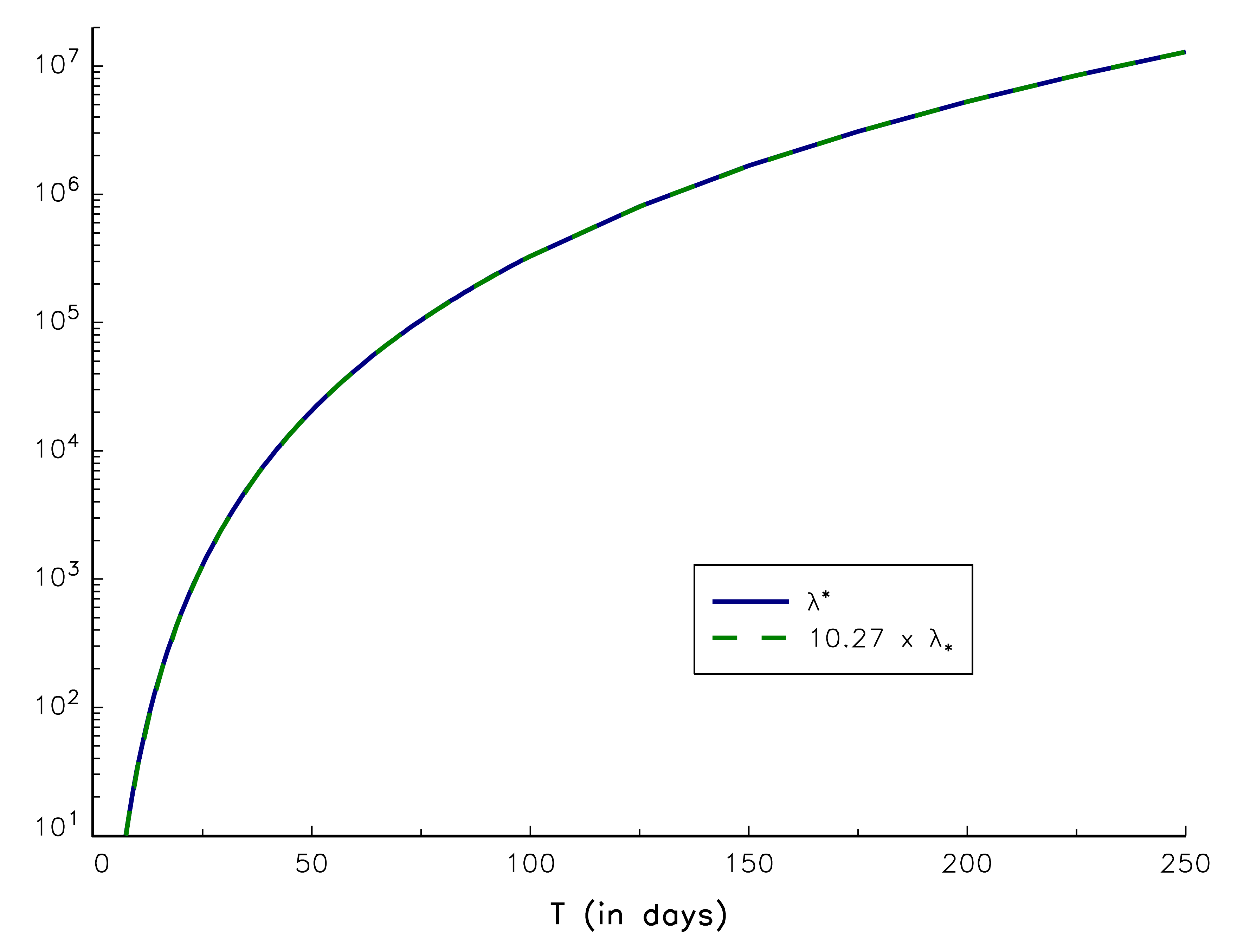}
\end{figure}

We discuss here how to calibrate the $L_{2}$ filter in order to extract the
trend with respect to the investment time horizon $T$. Though the $L_{2}$
filter admits an explicit solution which is a great advantage for numerical
implementation, the calibration of the smoothing parameter $\lambda $ is not
trivial. We propose to calibrate the $L_{2}$ filter by comparing the
spectral density of this filter with the one obtained with the
moving-average filter. For this last filter, we have:
\begin{equation*}
\hat{x}_{t}^{\mathrm{MA}}=\frac{1}{T}\sum_{i=t-T}^{t-1}y_{i}
\end{equation*}%
It comes that the spectral density is:%
\begin{equation*}
f\left( \omega \right) =\frac{1}{T^{2}}\left\vert
\sum_{t=0}^{T-1}e^{-i\omega t}\right\vert ^{2}
\end{equation*}%
For the $L_{2}$ filter, we k now that the solution is $\hat{x}^{\mathrm{HP}%
}=\left( 1+2\lambda D^{T}D\right) ^{-1}y$. Therefore, the spectral density
is:%
\begin{align*}
f^{\mathrm{HP}}\left( \omega \right) & =\left( \frac{1}{1+4\lambda \left(
3-4\cos \omega +\cos 2\omega \right) }\right) ^{2} \\
& \simeq \left( \frac{1}{1+2\lambda \omega ^{4}}\right) ^{2}
\end{align*}%
The width of the spectral density for the $L_{2}$ filter is then $\left(
2\lambda \right) ^{-1/4}$ whereas it is $2\pi T^{-1}$ for the moving-average
filter. Calibrate the $L_{2}$ filter could be done by matching this two
quantities. Finally, we obtain the following relationship:%
\begin{equation*}
\lambda \propto \lambda_\star = \frac{1}{2}\left( \frac{T}{2\pi
}\right) ^{4}
\end{equation*}%
In Figure \ref{fig:lambda4-a}, we represent the spectral density of
the moving-average filter for different windows $T$. We report also the
spectral density of the corresponding $L_2$ filters. For that, we have calibrated
the optimal parameter $\lambda^{\star}$ by least square minimization. In Figure
\ref{fig:lambda4-b}, we compare the optimal estimator $\lambda^{\star}$ with the one corresponding to
$10.27 \times \lambda_\star$. We notice that the approximation is very good.

\subsection{Implementation issues}
The computational time may be large when working with dense matrices even if
we consider interior-point algorithms.
It could be reduced by using sparse matrices. But the efficient way to
optimize the implementation is to consider band matrices. Moreover, we may
also notice that we have to solve a large linear system at each iteration.
Depending on the filtering problem ($L_1-T$, $L_1-C$ and $L_1-TC$ filters),
the system is 6-bands or 3-bands but always symmetric. For computing $\lambda_{\max}$,
one may remark that it is equivalent to solve a band system which is positive definite.
We suggest to adapt the algorithms in order to take into account all these properties.

\end{document}